\documentclass[pra,aps,twocolumn]{revtex4}
\usepackage{mathrsfs}
\usepackage{amsfonts}
\usepackage{txfonts}
\usepackage{amssymb}
\usepackage{graphicx}
\usepackage{bm}
\usepackage{color}

\newcommand{\ket}[1]{|#1\rangle}

\begin{document}
\title{Geometric phase of very slow neutrons}
\author{Erik Sj\"{o}qvist\footnote{Email: erik.sjoqvist@physics.uu.se}}
\affiliation{Department of Physics and Astronomy, Uppsala University, Box 516, 
Se-751 20 Uppsala, Sweden}
\begin{abstract} 
The geometric phase (GP) acquired by a neutron passing through a uniform magnetic field 
elucidates a subtle interplay between its spatial and spin degrees of freedom. In the 
standard setup using thermal neutrons, the kinetic energy is much larger than the typical 
Zeeman split. This causes the spin to undergo nearly perfect precession around the axis 
of the magnetic field and the GP becomes a function only of the corresponding cone angle. 
Here, we perform a plane wave analysis of the GP of very slow neutrons, for which the 
precession feature breaks down. Purely quantum-mechanical matter wave effects, such 
as resonance, reflection, and tunneling, become relevant for the behavior of the GP in this 
low energy scattering regime. 
\end{abstract}
\maketitle 
\section{Introduction}
When a spin-polarized neutron passes through a uniform magnetic field, it acquires 
a measurable phase that depends only on the path taken by the spin. This geometric
phase (GP)  is traditionally calculated under the assumption that the neutron moves at sufficiently 
large speed so that the effect of the magnetic field can be treated as a small perturbation 
to the free motion \cite{bernstein67}. In this `high-speed' case, the spin performs pure precession 
around the direction of the magnetic field; for one such precession cycle, the GP  
becomes \cite{aharonov87} 
\begin{eqnarray}
\gamma = -\pi (1-\cos \theta) , 
\label{eq:cyclicgp}
\end{eqnarray}
where $\theta$ is the angle between the spin and the magnetic field. This result 
has been confirmed and examined experimentally in the past 
\cite{wagh97,hasegawa01,klepp08}. 

Here, we ask: how is the GP modified if the neutron moves at very low 
speed, i.e., when its kinetic energy is comparable to the Zeeman split? In this case, 
the magnetic field cannot be treated as a small perturbation and the neutron spin no 
longer performs pure precession. The situation must be treated as a scattering problem, 
in which the spin components along the direction of the magnetic field are scattered 
differently, causing a speed dependent spin path. 

We perform a plane wave analysis of the spin dependent scattering of the neutron through 
the magnetic field region. We derive results for the GP in different energy regimes and discuss 
the feasibility of experimental parameters needed to see measurable deviations from the ideal 
GP in Eq.~(\ref{eq:cyclicgp}). We examine the influence of resonance, reflection, and tunneling 
effects on the GP; purely quantum-mechanical matter wave effects that become relevant in 
this low energy scattering regime. 

\section{Model}

\begin{figure}[h]
\begin{center}
\includegraphics[width=12 cm]{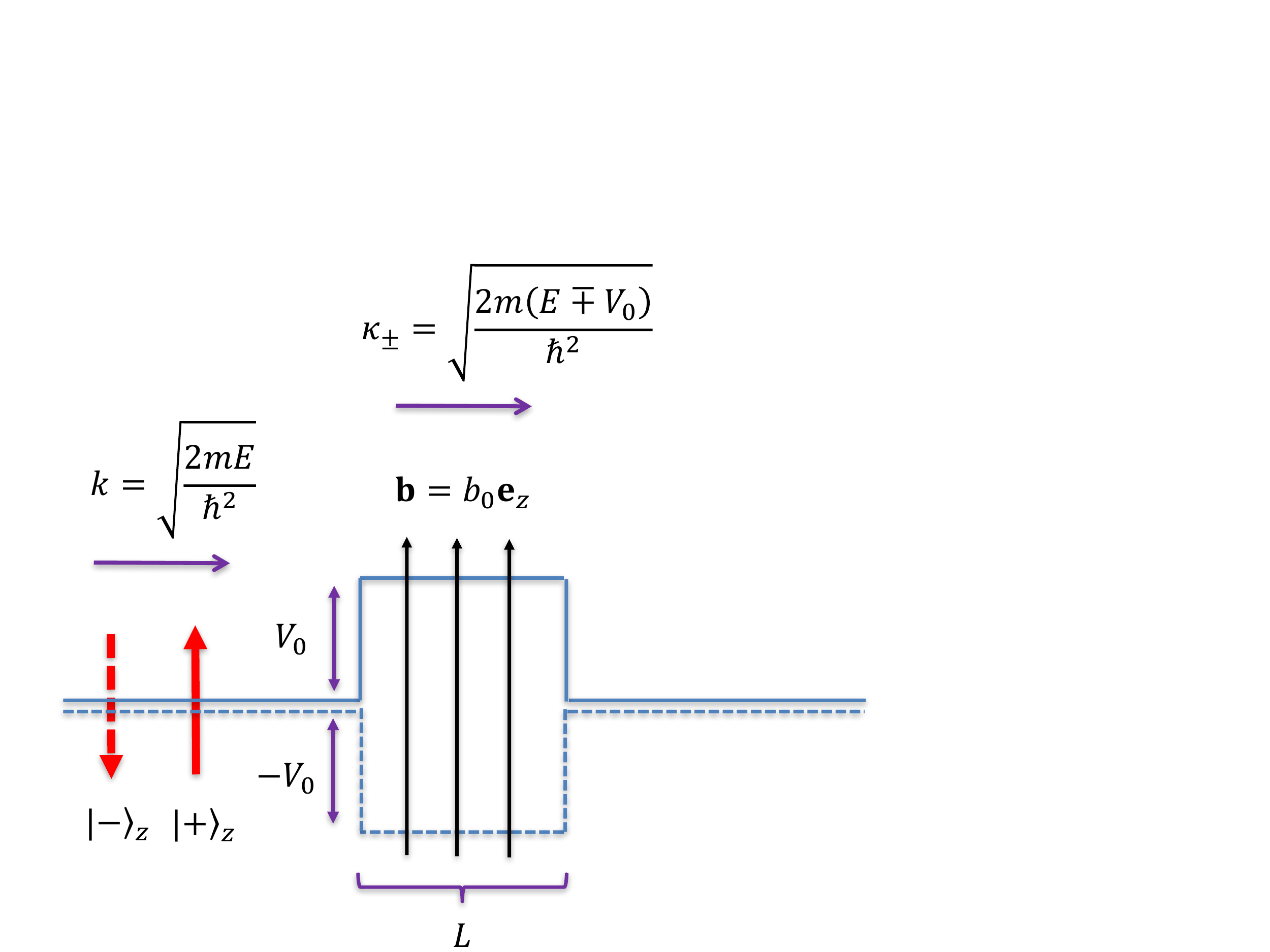}
\end{center}
\caption{(Color online) Setup for realizing a nontrivial GP. Neutrons with energy $E$ 
enter a homogenous magnetic field region of width $L$. The spin up (down) sees 
a potential barrier (well) due to the Zeeman interaction. This results in the wave vectors  
$\kappa_{\pm}$ for the two spin components in the magnetic field region. Here, 
$2V_0 = \mu b_0$ is the Zeeman split, $\mu$ being the magnetic moment of the neutron. 
The difference $\kappa_{+} - \kappa_{-}$ causes spin precession and a nontrivial GP.}
\label{fig:basic_setup}
\end{figure}

The basic setup is shown in Fig.~\ref{fig:basic_setup}. The linear motion of the neutron is 
described by the Hamiltonian 
\begin{eqnarray}
H = -\frac{\hbar^2}{2m} \frac{d^2}{dx^2} + \boldsymbol{\mu} \cdot {\bf b} . 
\end{eqnarray}
Here, $m=1.675 \times 10^{-27}$ kg and $\boldsymbol{\mu} = \mu \frac{1}{2} \boldsymbol{\sigma}$ 
with $\mu = 9.662 \times 10^{-27}$ J/T are the mass and magnetic moment, respectively, 
of the neutron. $\boldsymbol{\sigma} = \left( \sigma_x,\sigma_y,\sigma_z \right)$ is the 
vector of the standard Pauli operators.  We further introduce the spin basis $\ket{\pm}_z$ 
such that $\sigma_z \ket{\pm}_z = \pm \ket{\pm}_z$. The magnetic field is taken to be  
\begin{eqnarray}
{\bf b} = \left\{ \begin{array}{ll}  
b_0 {\bf e}_z, & 0 \leq x \leq L, \\ 
0, & {\rm otherwise,} 
\end{array} \right. 
\end{eqnarray}
which implies a spin dependent potential $V(x) = V_0 \sigma_z $, $0 \leq x \leq L$, and 
$V(x) = 0$ otherwise, where $2V_0 = \mu b_0$ is the Zeeman split between 
the two spin states $\ket{\pm}_z$. The coordinate system is chosen so that $V_0 > 0$. 
We assume the kinetic energy of the incoming neutron is $E$. The behavior of the 
GP is essentially determined by the ratio 
\begin{eqnarray}
\frac{V_0}{E} = \frac{\mu b_0}{mv^2} ,  
\label{eq:ideality}
\end{eqnarray}
the ideal case being $\frac{V_0}{E} \ll 1$. For example, a typical triple-Laue experiment 
utilizes neutrons with speed $v \sim 2000$ m/s and magnetic fields $b_0 \sim 10^{-2}$ T, 
yielding a ratio $\frac{V_0}{E} \sim 10^{-8}$, which is well within this ideal regime. 

The spin evolution can be treated as a one-dimensional scattering problem. Let 
\begin{eqnarray}
\ket{\psi} = c_{+} \ket{+}_z + c_{-} \ket{-}_z & = & \sum_{l=\pm} c_{l} \ket{l}_z, 
\nonumber \\   
\left| c_{+} \right|^2 + \left| c_{-} \right|^2 & = & 1, 
\end{eqnarray}
be the normalized spin state at $x=0$, which is where the neutron enters the magnetic 
field region. While the $\ket{+}_z$ state sees the potential barrier $V_0$ when entering 
the region, the $\ket{-}_z$ state sees the potential well $-V_0$. We can solve for these 
two cases in terms of the wave numbers 
\begin{eqnarray}
\kappa_{\pm} = \sqrt{\frac{2m(E\mp V_0)}{\hbar^2}} \equiv \sqrt{k^2 \mp \kappa_{0}^2} , 
\label{eq:effective_wl}
\end{eqnarray}
$k$ and $\kappa_{0}$ being determined by the speed of the incoming neutron and the 
Zeeman split between its two spin states, respectively. We can write the spatial wave 
function for $x\leq 0$ as $\psi_{\pm} (x) = 
e^{ikx}+\sqrt{r_{\pm}} e^{i\delta_{\pm}} e^{-ikx}$ with $r_{\pm}$ the reflection probability and 
$\delta_{\pm}$ the scattering phase shift acquired upon reflection. We further write 
$\psi_{\pm} (x) = A_{\pm} e^{i\kappa_{\pm} x} + B_{\pm} e^{-i\kappa_{\pm} x}$ and 
$\psi_{\pm} (x) = C_{\pm} e^{ik(x-L)}$ on $0\leq x \leq L$ and $x\geq L$, respectively. 
The matching conditions at $x=0$ and $x=L$ yield 
\begin{eqnarray}
\sqrt{r_{\pm}} e^{i\delta_{\pm}} & = & 
\frac{(k^2 - \kappa_{\pm}^2) \sin \kappa_{\pm} L}{ (k^2 + \kappa_{\pm}^2) \sin \kappa_{\pm} L +
2ik\kappa_{\pm} 
\cos \kappa_{\pm} L} 
\label{eq:reflection}
\end{eqnarray}
and 
\begin{eqnarray}
\frac{B_{\pm}}{A_{\pm}} & = & 
e^{2i\kappa_{\pm}L} \left( \frac{\kappa_{\pm} - k}{\kappa_{\pm} + k} \right) . 
\end{eqnarray}
The neutron spin can be followed through the setup:  
\begin{itemize} 
\item[(i)] $x \leq 0$: 
\begin{eqnarray}
\ket{\phi^{\rm (i)} (x)} & = & \sum_{l=\pm}
\left( \frac{e^{ikx}+\sqrt{r_{l}} e^{i\delta_{l}} e^{-ikx}}{1+\sqrt{r_{l}} e^{i\delta_{l}}} \right) c_{l} \ket{l}_z , 
\label{eq:wfi}
\end{eqnarray}
where we have normalized the spin components so as to satisfy $\ket{\phi^{\rm (i)} (0)} = \ket{\psi}$. 
\item[(ii)] $0 \leq x \leq L$:  
\begin{eqnarray}
 & & \ket{\phi^{\rm (ii)} (x)} 
\nonumber \\ 
 & & = \sum_{l=\pm} \left( \frac{\kappa_{l} \cos \left[ \kappa_{l} (x-L) \right] + 
i k\sin \left[ \kappa_{l} (x-L) \right]}{\kappa_{l} \cos \kappa_{l} L - 
i k\sin \kappa_{l} L} \right) c_{l} \ket{l}_z , 
\label{eq:wfii}
\end{eqnarray}
again with $ \ket{\phi^{\rm (ii)} (0)} = \ket{\psi}$. 
\item[(iii)] $x \geq L$:
\begin{eqnarray}
 \ket{\phi^{\rm (iii)} (x)} = e^{ik(x-L)} \sum_{l=\pm} \frac{\kappa_{l}}{\kappa_{l} \cos \kappa_{l} L - 
i k\sin \kappa_{l} L} c_{l} \ket{l}_z . 
\label{eq:wfiii}
\end{eqnarray}
Since the $x$ dependence is fully confined to the overall phase factor $e^{ik(x-L)}$, the 
GP vanishes in this region for all choices of Zeeman split and speed of the incoming neutron. 
\end{itemize} 

To cover both cyclic and noncyclic evolution of the neutron spin, we use the open 
path GP \cite{mukunda93}. This phase takes the form 
\begin{widetext}
\begin{eqnarray}
\gamma & = & \arg \langle \phi (x_0) \ket{\phi (x_1)} + 
\frac{i}{2} \int_{x_0}^{x_1} \left( \frac{\langle \phi (x) \ket{\frac{d}{dx} \phi (x)} - 
\langle \frac{d}{dx} \phi (x) \ket{\phi (x)}}{\langle \phi (x) \ket{\phi (x)}} \right) dx 
\nonumber \\ 
 & = & \arg \left[ \sum_{l=\pm} f_{l}^{\ast} (x_0) f_{l} (x_1) \right] + 
\frac{i}{2} \int_{x_0}^{x_1} \left[ \sum_{l=\pm} \left| f_l (x) \right|^2 \right]^{-1}
\sum_{l=\pm} \left[ f_{l}^{\ast} (x) \frac{d}{dx} f_{l} (x) - f_{l} (x) \frac{d}{dx} f_{l}^{\ast} (x) \right] dx , 
\label{eq:gengp}
\end{eqnarray}
\end{widetext}
where we have introduced the $x$ dependent complex-valued amplitudes $f_{l} (x)$ 
according to 
\begin{eqnarray}
\ket{\phi (x)} = \sum_{l=\pm} f_{l} (x) \ket{l}_z .  
\end{eqnarray}
Equation (\ref{eq:gengp}) can be applied to any interval $x_0 \leq x \leq x_1$ by identifying 
$f_{l} (x)$ from Eqs.~(\ref{eq:wfi})-(\ref{eq:wfiii}).  

\section{Cyclic GP in resonant scattering}
The energy barrier becomes transparent if the length $L$ is an integer multiple of half 
the effective wave length of the neutron inside the magnetic field region. If this resonance 
condition is satisfied for both spin state components simultaneously, the spin undergoes 
loss-free unitary evolution through the magnetic field. If the winding numbers of phase factors of 
the two spin states are both even or both odd, the spin performs cyclic evolution. This 
creates a GP of cyclic evolution enforced by the simultaneous resonance scattering of 
the two spin states; this low-energy GP can be compared to the ideal GP in 
Eq.~(\ref{eq:cyclicgp}). 

As is evident from Eq.~(\ref{eq:wfii}), the probability amplitudes undergo the change 
\begin{eqnarray}
c_{l} \mapsto \frac{\kappa_{l}}{\kappa_{l} \cos \kappa_{l} L - 
i k\sin \kappa_{l} L} c_{l} 
\end{eqnarray}
during the passage through the magnetic field. Thus, the resonance conditions 
that enforce cyclic evolution become 
\begin{eqnarray}
\kappa_{l} = \frac{\pi}{L} n_{l} , 
\label{eq:resonanceconditions}
\end{eqnarray}
where $n_{l}$ are either both even or both odd integers \cite{remark}. Note that these conditions 
can only be satisfied if $E > V_0$. Furthermore, since $\kappa_{-} > \kappa_{+}$ for $V_0 > 0$ 
it follows that $n_{-} > n_{+}$. The resonance conditions put restrictions on the Zeeman split 
and the speed of the incoming neutron according to  
\begin{eqnarray}
\kappa_0  & = & \frac{\pi}{L} \sqrt{\frac{n_{-}^2 - n_{+}^2}{2}} , 
\nonumber \\ 
k & = & \frac{\pi}{L} \sqrt{\frac{n_{-}^2 + n_{+}^2}{2}} . 
\label{eq:resonance}
\end{eqnarray}
As can be seen from Eq.~(\ref{eq:reflection}), the conditions in Eq.~(\ref{eq:resonanceconditions}) 
imply $r_{l} = 0$. Thus, the GP $\gamma^{\rm (i)}$ on $x \leq 0$ vanishes 
and a nontrivial GP can only be acquired on $0 \leq x \leq L$ in this case. Equation (\ref{eq:wfii}) 
reduces to 
\begin{eqnarray}
\ket{\phi^{\rm (ii)} (s)} = \sum_{l=\pm} \left( \cos n_{l} s + 
i \frac{k}{\kappa_{l}} \sin n_{l} s \right) c _{l} \ket{l}_z , 
\end{eqnarray}
where we have defined $s=(\pi/L)x$ such that $s\in [0,\pi]$ over the magnetic field region. Since 
$n_{l}$ are either both even or both odd and $n_{-} > n_{+}$, we can write $n_{-} = n_{+} + 2\xi$, 
$\xi$ being a positive integer. It follows that  
\begin{eqnarray}
\ket{\phi^{\rm (ii)} (\pi)} = (-1)^{n_{+}} \ket{\phi^{\rm (ii)} (0)}. 
\label{eq:bc}
\end{eqnarray}

In the high-speed limit where $k\gg \kappa_0$ \cite{bernstein67}, we have 
$\kappa_{l} \approx k$, which implies 
\begin{eqnarray}
\ket{\phi^{\rm (ii)} (s)} \approx \sum_{l=\pm} e^{in_{l} s} c _{l} \ket{l}_z . 
\end{eqnarray}
The corresponding GP becomes 
\begin{eqnarray}
\gamma^{\rm (ii)} & \approx & \arg \langle \phi^{\rm (ii)} (0) \ket{\phi^{\rm (ii)} (\pi)} + 
i \int_0^{\pi} \langle \phi^{\rm (ii)} (s) \ket{\frac{d}{dx} \phi^{\rm (ii)} (s)} ds 
\nonumber \\ 
 & = & \pi n_{+} - \pi \left( n_{-} \left| c_{-} \right|^2 + n_{+} \left| c_{+} \right|^2  \right) =  
- \xi 2\pi \left| c_{-} \right|^2 .  
\label{eq:approxgp1}
\end{eqnarray}
By letting $\left| c_{-} \right| = \sin \frac{\theta}{2}$, $\theta$ being the angle between the 
spin and the magnetic field, we may rewrite Eq.~(\ref{eq:approxgp1}) as 
\begin{eqnarray}
\gamma^{\rm (ii)} & \approx & - \xi \pi ( 1-\cos \theta ) , 
\label{eq:largevgp}
\end{eqnarray}
which coincides with the expected expression in Eq.~(\ref{eq:cyclicgp})  up to the 
`winding number' $\xi$ that counts the number of precessions of the spin when 
passing through the magnetic field. 

Now, by turning to the exact treatment, we use Eq.~(\ref{eq:gengp}) and find 
\begin{eqnarray}
\gamma^{{\rm (ii)}} & = & \pi n_{+} - \sqrt{\frac{n_{-}^2 + n_{+}^2}{2}} 
 \nonumber \\ 
 & & \times \int_0^{\pi} 
 \left\{ \sum_{l=\pm} 
\left( \cos^2 n_{l} s + \frac{n_{-}^2 + n_{+}^2}{2 n_{l}^2} \sin^2 n_{l} s \right) \left| c_{l} \right|^2  
\right\}^{-1} ds.
\nonumber \\ 
\label{eq:exactgp}
\end{eqnarray}
This expression is valid for any parameter choice as long as $k > \kappa_0$ and the 
resonance conditions in Eq.~(\ref{eq:resonanceconditions}) are satisfied. 

\begin{figure}[h]
\begin{center}
\includegraphics[width=7.5 cm]{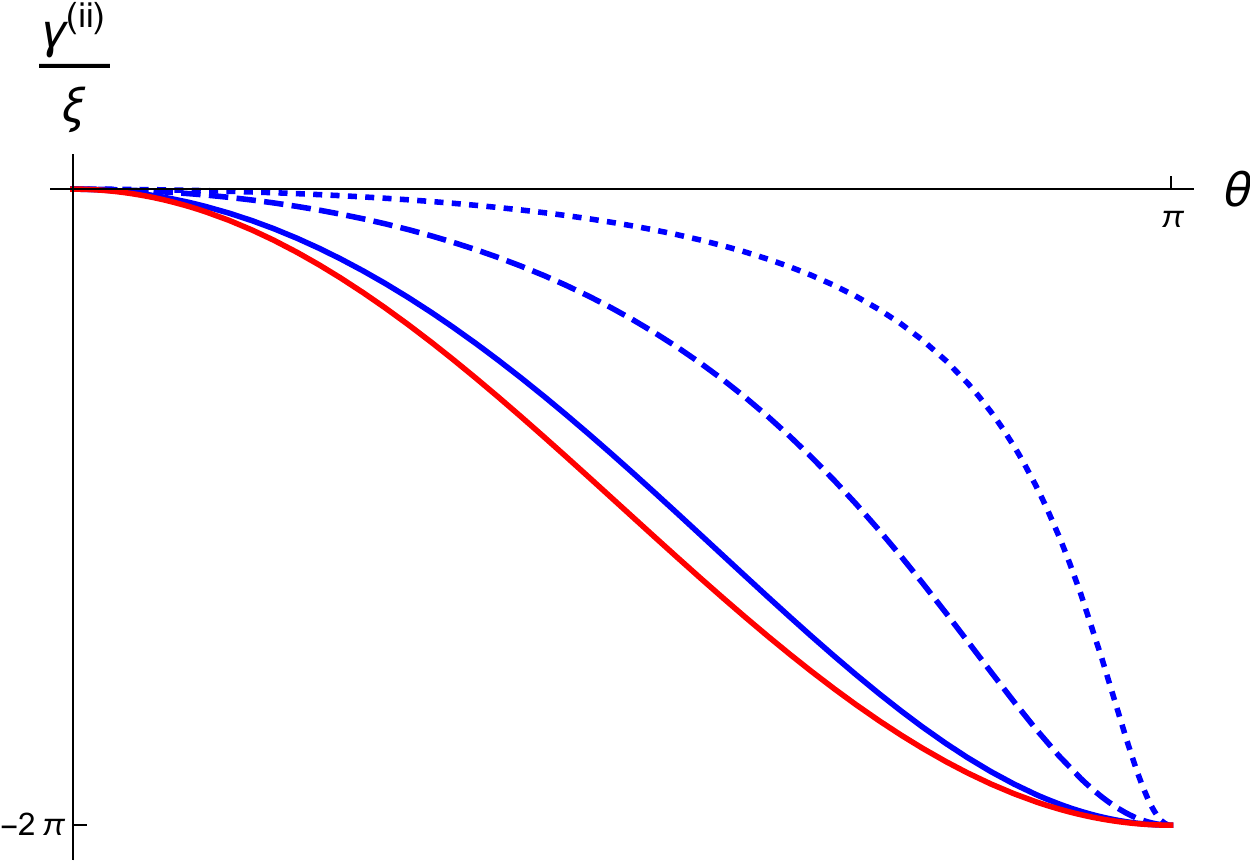}
\end{center}
\caption{(Color online) GP per turn $\frac{\gamma^{{\rm (ii)}}}{\xi}$  as a function of 
the angle $\theta$ between the magnetic field and the incoming spin. The solid, dashed, and 
dotted blue lines correspond to $\frac{n_{+}}{n_{-}} = \frac{1}{q}$ with $q=1.2,3,$ and $10$, 
respectively. The red line shows the reference ideal GP $-\pi (1-\cos \theta)$ 
obtained in the $q \rightarrow 1$ limit.}
\label{fig:gpresonance}
\end{figure}

We examine numerically the deviation of the GP per turn $\frac{\gamma^{\rm (ii)}}{\xi}$ 
in terms of the ratio $\frac{n_{+}}{n_{-}} \equiv \frac{1}{q}$, $q>1$. Figure \ref{fig:gpresonance} 
shows the exact GP per turn for $q=1.2,3,$ and $10$, as the solid, dashed, and dotted 
blue lines, respectively. We find a considerable deviation from the ideal GP (shown as 
the red line) obtained in the $q \rightarrow 1$ limit. Further note that $\gamma^{\rm (ii)}$ 
vanishes (mod $2\pi$) for $\theta = 0$ (input state $\ket{+}_z$) or $\pi$ (input state 
$\ket{-}_z$); a feature that can be seen directly from Eq.~(\ref{eq:wfii}) by putting either 
$c_{-}=0$ or $c_{+}=0$. 

By combining Eqs.~(\ref{eq:ideality}), (\ref{eq:effective_wl}), and (\ref{eq:resonance}), we 
can relate the magnetic field strength $b_0$, the neutron speed $v$, and the parameter 
$q$ according to 
\begin{eqnarray}
b_0 [{\rm T}] & = & \left( \frac{q^2-1}{q^2+1} \right) 0.17 \times 
\left( v[{\rm ms}^{-1}] \right)^2 . 
\label{eq:bvrelation}
\end{eqnarray}
We see that thermal neutrons, corresponding to $v \sim 2000$ ms$^{-1}$, would 
require $q^2 \sim 1 + 10^{-7}$ in order to achieve a manageable magnetic field strength, 
say, about $1$ T. For such small $q$, $\gamma^{\rm (ii)}$ is practically indistinguishable  
from the ideal GP. We need instead to consider ultracold neutrons, typically 
having speed in the order of $1$ ms$^{-1}$, to achieve a measurable 
difference between the exact and ideal GPs for feasible magnetic 
field strengths. Indeed, we find $b_0 = 0.17 \ {\rm T}, 0.14 \ {\rm T}, 0.031 \ {\rm T}$ 
for $q=10, 3, 1.2$, respectively, when $v = 1$ m/s. 

\section{GP in off-resonant case}
In the off-resonant case, the low-energy GP shows new phenomena since the transmission 
probabilities of the two spin states are no longer unity. This means that the neutron beam 
is partly backscattered by the magnetic field. 

\subsection{GP for $x < 0$}
A key novel aspect in the off-resonance case is that the spin state may be $x$ dependent 
for $x < 0$. This implies that the neutron may pick up a nontrivial GP even 
before it enters the magnetic field. 

\begin{figure}[ht]
\begin{center}
\includegraphics[width=7.5 cm]{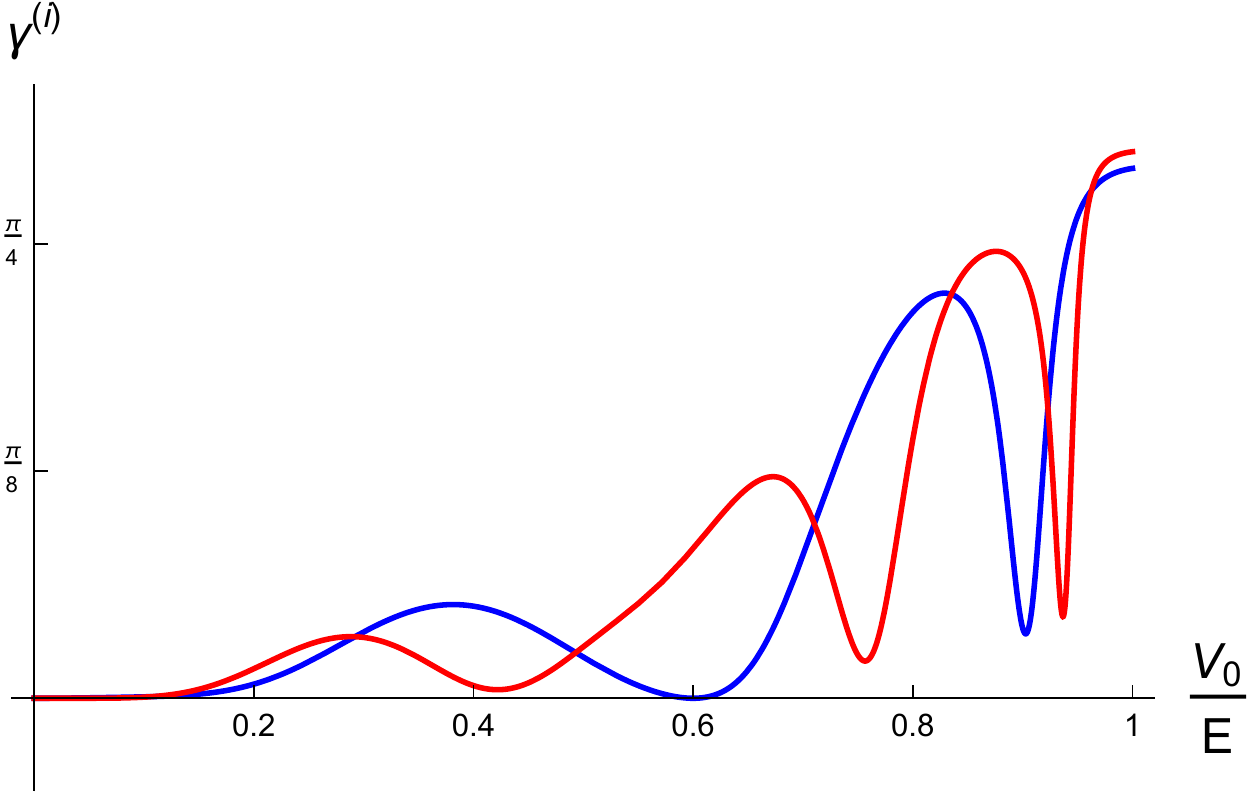}
\end{center}
\caption{(Color online) GP $\gamma^{\rm (i)}$ for one cycle as a function of 
$\frac{V_0}{E}$ for the symmetric input state $c_{-} = c_{+} = \frac{1}{\sqrt{2}}$. 
The blue curve shows the GP for $kL=\sqrt{10} \pi$, which admits resonance 
scattering at $\frac{V_0}{E} = \frac{3}{5}$, corresponding to $n_{-} = 4$ and $n_{+} = 2$. 
At this energy ratio, the magnetic field becomes fully transparent for both spin components  
and $\gamma^{\rm (i)}$ would vanish, as confirmed by the simulation. The red curve shows 
$kL = 4\pi$, for which no nonzero energy ratio can give rise to $r_{\pm} = 0$ simultaneously. 
A closer inspection shows that $\gamma^{\rm (i)}$ is indeed nonvanishing for all $\frac{V_0}{E} > 0$.}
\label{fig:gpnegativex}
\end{figure}

The spin state given by Eq.~(\ref{eq:wfi}) repeatedly undergoes cyclic evolution with period 
$\pi/k$ for any starting point $x_0 \leq -\pi/k$. It follows that the GP 
$\gamma_{\zeta}^{\rm (i)}$ for any number $\zeta$ of cycles is independent of $x_0$ and can 
be written as $\gamma_{\zeta}^{\rm (i)} = \zeta \gamma^{\rm (i)}$, where $\gamma^{\rm (i)}$ is the GP 
for one cycle. We find 
\begin{eqnarray}
 & & \gamma^{\rm (i)} = \pi - \sum_{l=\pm} \left( \frac{1-r_{l}}{1 + r_{l} + 2\sqrt{r_{l}} \cos \delta_{l}} \right) 
\left| c_{l} \right|^2 
\nonumber \\ 
 & & \times \int_0^{\pi} \left\{ \sum_{l=\pm} 
\left[ \frac{1 + r_{l} + 2\sqrt{r_{l}} \cos \left( \delta_{l} - 2s \right)}{1+ r_l + 
2\sqrt{r_{l}} \cos \delta_{l}} \right]  \left| c_{l} \right|^2 \right\}^{-1} ds,
\nonumber \\ 
\label{eq:negativexgp}
\end{eqnarray}
where the integral is expressed in terms of the dimensionless variable $s=kx$ and we have 
made use of the $x_0$ independence to shift the integration interval to $[0,\pi]$. Thus, the 
neutron may acquire a nonzero GP before it enters the magnetic 
field region, due to a nonzero reflection probability of at least one of the two spin states . 

The GP $\gamma^{\rm (i)}$ is fully determined by the energy ratio $\frac{V_0}{E}$ 
and the kinematic parameter $kL$ via the reflection probabilities $r_{l}$ and phase shifts 
$\delta_{l}$. In Fig.~\ref{fig:gpnegativex}, we show $\gamma^{\rm (i)}$ as a function of $\frac{V_0}{E}$ 
for two different $kL$ and the symmetric input state $c_{-} = c_{+} = \frac{1}{\sqrt{2}}$. We have 
chosen $kL = \sqrt{10} \pi$ (blue line), for which there is an 
energy ratio that corresponds to resonant scattering and vanishing $\gamma^{\rm (i)}$. Explicitly, we 
find this ratio to be $\frac{V_0}{E} = \frac{3}{5}$ with $n_{-} = 4$ and $n_{+} = 2$. We confirm 
in Fig.~\ref{fig:gpnegativex} that $\gamma^{\rm (i)}$ vanishes at this energy ratio. The other 
choice (red line) is $kL = 4\pi$ with resonance solution $n_{-} = n_{+} = 4$ corresponding to the 
trivial case $V_0 = 0$. Thus, the barrier never becomes transparent for both spin components 
simultaneously for this choice, and the GP $\gamma^{\rm (i)}$ is nonvanishing for 
all $\frac{V_0}{E} > 0$, as is confirmed by the simulation.

The incoming neutron speed corresponding to the two curves in Fig.~\ref{fig:gpnegativex} 
are very low and have been chosen mainly for illustrative purposes. Indeed, for 
$kL \sim 10$, one obtains $v[{\rm m/s}] \sim 10^{-6} / L[{\rm m}]$, which would imply a 
reasonable high speed only for very small $L$. In principle, this could be achieved by 
using a thin metallic foil to create the magnetic field, for instance by using a setup 
similar to that of Ref.~\cite{klein76}.  

\begin{figure}[ht]
\begin{center}
\includegraphics[width=7.5 cm]{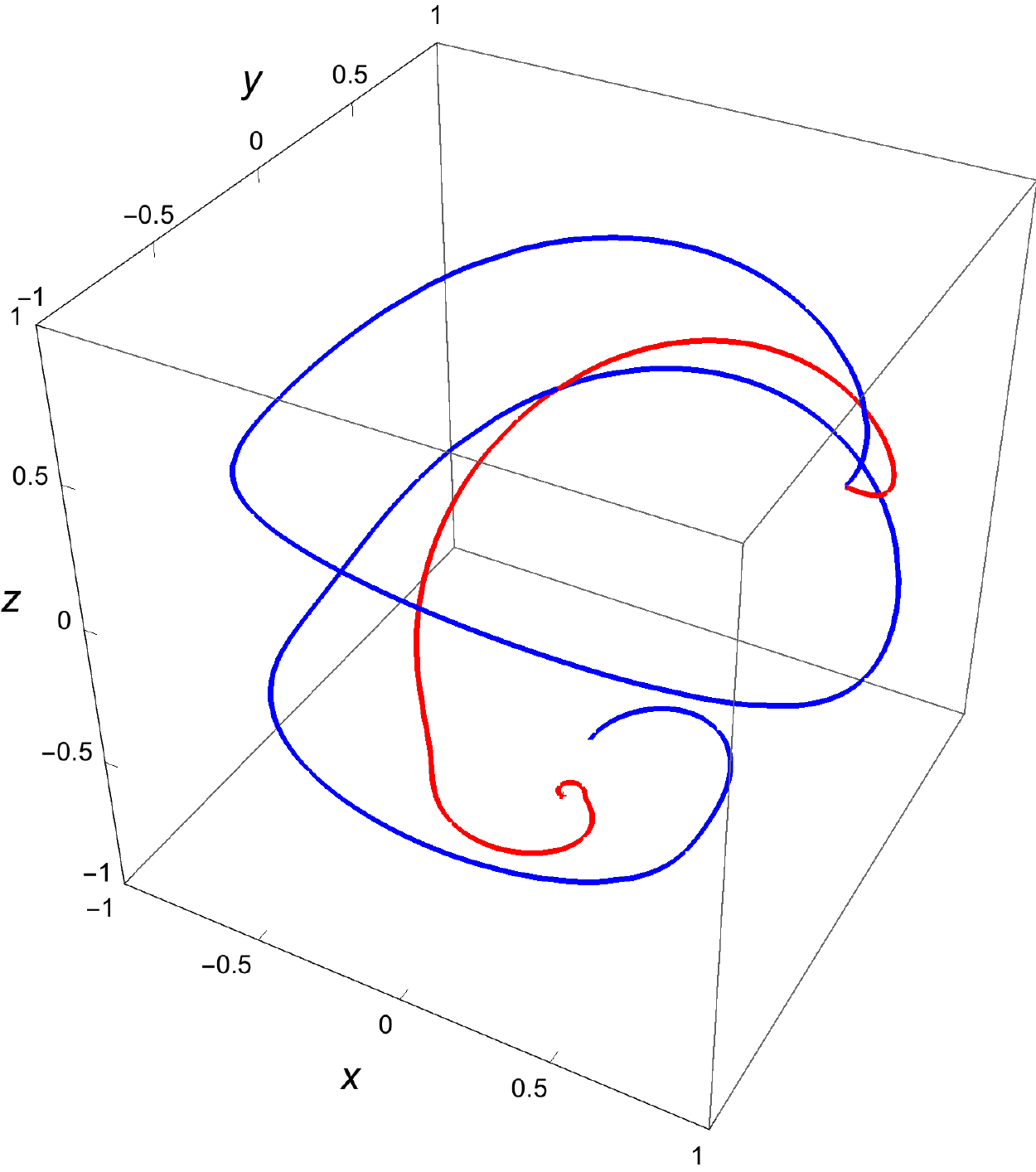}
\end{center}
\caption{(Color online) Evolution of the spin direction when the neutron passes through 
the magnetic field. The wave vector $\kappa_{-}$ of the $\ket{-}_z$ spin state satisfies 
$\kappa_{-} L = 5\pi$. We have chosen $\frac{V_0}{E} = 1.01$ (blue curve) and 
$\frac{V_0}{E} = 2$ (red curve), both starting at $(c_{+},c_{-}) = (\frac{\sqrt{3}}{2},\frac{1}{2})$. 
The evolution is genuinely noncyclic. The decay towards the $\ket{-}_z$ state at the negative 
$z$ axis is clearly visible.}
\label{fig:blochvector}
\end{figure}

\subsection{Tunneling-induced GP}
For extremely slow neutrons, at kinetic energies corresponding to $k < \kappa_0$, the 
classical intuition entails that the magnetic field would act as a filter that allows only the 
spin state $\ket{-}_z$ to enter the magnetic field region, in case of which $\gamma^{\rm (ii)}$ 
would be expected to vanish. However, a portion of the $\ket{+}_z$ state is transmitted via 
the tunnel effect. This creates a tunneling-induced GP. 

While the  $\ket{-}_z$ state undergoes an oscillatory evolution, the $\ket{+}_z$ state decays 
during the passage through the magnetic field when $k < \kappa_0$. In this regime, we may 
write $\kappa_{+} = i \kappa$ with $\kappa = \sqrt{\kappa_0^2 - k^2}$, yielding  
\begin{eqnarray}
c_{+} \mapsto \left( \frac{\kappa \cosh \left[ \kappa (x-L)\right] + 
ik \sinh \left[ \kappa (x-L)\right]}{\kappa 
\cosh \kappa L - ik \sinh \kappa L} \right) c_{+} . 
\end{eqnarray}
Figure \ref{fig:blochvector} shows the evolution of the spin direction for the entire 
passage of the neutron through the magnetic field with $\kappa_{-} L = 5 \pi$ and $(c_{+},c_{-}) = 
(\frac{\sqrt{3}}{2},\frac{1}{2})$. We choose $\frac{V_0}{E} = 1.01$ and $2$ 
corresponding to, respectively, an energy of the incoming neutron slightly below and 
half the magnetic-field-induced barrier seen by the $\ket{+}_z$ state. The curves can 
be understood as a combined decay and oscillatory motion. The spin spirals down 
to the $\ket{-}_z$ state, pointing in the negative $z$ direction. The evolution is genuinely 
noncyclic. 

\begin{figure}[ht]
\begin{center}
\includegraphics[width=7.5 cm]{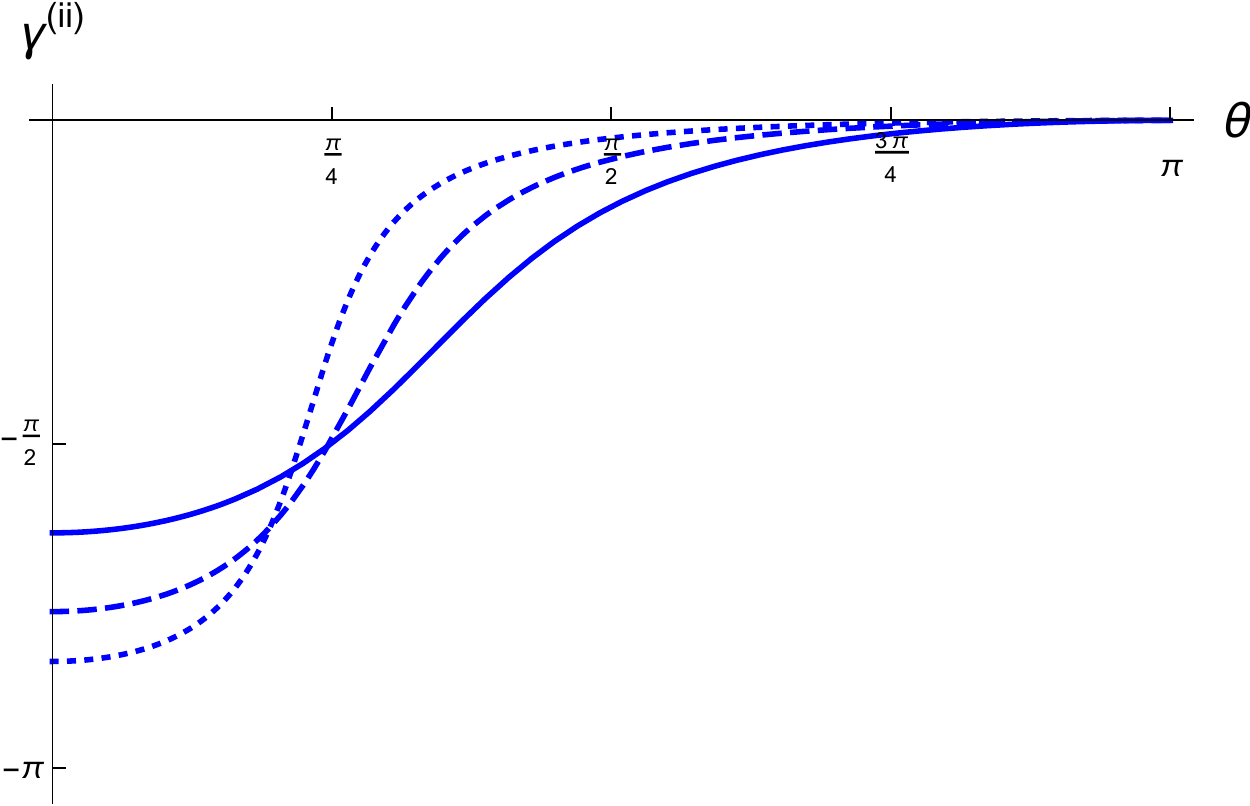}
\end{center}
\caption{(Color online) Tunneling-induced GP as a function of the angle $\theta$ 
between the directions of the input spin and the magnetic field. The wave vector $\kappa_{-}$ 
of the $\ket{-}_z$ spin state satisfies $\kappa_{-} L = \pi$. The solid, dashed, and dotted 
lines show $\frac{V_0}{E} = 1.01, 2,$ and $5$, respectively.}
\label{fig:tunnelgp}
\end{figure}

Figure \ref{fig:tunnelgp} shows the tunneling-induced GP as a function 
of the angle $\theta$ between the incoming spin and the magnetic field. We see that 
the GP is quenched more rapidly when the kinetic energy tends to zero. 
It remains large, however, for initial spins close to the positive $z$ axis, since the 
spin is forced to traverse a long path towards the negative $z$ direction, thereby 
enclosing a substantial solid angle.  

\section{Conclusions}
Spin-polarized neutrons passing through a uniform magnetic field have offered a rich 
field to study subtle aspects of the geometric phase (GP) experimentally. These studies have 
typically used neutrons moving at sufficiently high speed so that the magnetic field can be 
treated as a small perturbation to the free motion. In this work, we relax the high energy 
requirement and consider the case where the kinetic energy of the incoming neutron is 
comparable to the Zeeman shift. In this regime, the two spin states along the direction 
of the magnetic field scatter in such a way that the spin evolution and the concomitant 
GP deviate significantly from the pure precession that occurs at high energies.  

We have demonstrated the existence of cyclic spin evolution during the passage through 
the magnetic field for low energy neutrons and have calculated the corresponding GP. 
The evolution becomes cyclic if the two spin components simultaneously satisfy 
certain resonance conditions. When these resonance conditions do not hold, however, 
the incoming spin state may become nontrivially $x$ dependent and may pick up a nonzero 
GP even before the neutron enters the magnetic field.  
At even lower velocities, one eventually enters the tunneling regime, where one of the 
spin components decays while the other remains in an oscillatory mode. In this case, 
we have shown the existence of a tunneling-induced GP that can be large, even 
when the kinetic energy of the incoming neutron is smaller than the Zeeman shift. 

The realization of the GP effects discussed in this work requires a combination 
of low velocity and high magnetic field that put restrictions on the experimental feasibity. 
For instance, thermal neutrons, moving at speed in the order of $2000$ ms$^{-1}$, would 
require unrealistically large magnetic fields to see deviations from the ideal case. Therefore, 
to verify experimentally the rich behavior of the GP, ultracold neutrons moving 
at speed in the order of a few ms$^{-1}$ would be needed. 

Another possibility is to replace the neutrons with atoms, for 
which the magnetic moment is larger, still having a mass of the same order of magnitude. 
For instance, for atomic hydrogen the magnetic moment would be dominated 
by that of the single electron, which is essentially the Bohr magneton $\mu_B = 9.27 
\cdot 10^{-24}$ J/T. For this atom, one would therefore expect the feasible velocity 
range to be increased by a factor of $\left( \mu_B / \mu \right)^{1/2} \sim 30$ in order to see 
the low-energy GP effects.  

\section*{Acknowledgments}
Financial support from the Swedish Research Council (VR) through Grant No. 
2017-03832 is acknowledged.


\begin{thebibliography}{99}
\bibitem{bernstein67} H. J. Bernstein, 
Spin Precession During Interferometry of Fermions and the Phase Factor Associated with 
Rotations Through $2\pi$ Radians, 
Phys. Rev. Lett. {\bf 18}, 1102 (1967). 
\bibitem{aharonov87} Y. Aharonov and J. Anandan, 
Phase change during a cyclic quantum evolution, 
Phys. Rev. Lett. {\bf 58}, 1593 (1987). 
\bibitem{wagh97} A. G. Wagh, V. C. Rakhecha, J. Summhammer, G. Badurek, H. Weinfurter, 
B. E. Allman, H. Kaiser, K. Hamacher, D. L. Jacobson, and S. A. Werner, 
Experimental Separation of Geometric and Dynamical Phases Using Neutron Interferometry, 
Phys. Rev. Lett. {\bf 78}, 755 (1997). 
\bibitem{hasegawa01} Y. Hasegawa, R. Loidl, M. Baron, G. Badurek, and H. Rauch, 
Off-Diagonal Geometric Phase in a Neutron Interferometer Experiment, 
Phys. Rev. Lett. {\bf 87}, 070401 (2001).  
\bibitem{klepp08} J. Klepp, S. Sponar, S. Filipp, M. Lettner, G. Badurek, and Y. Hasegawa, 
Observation of Nonadditive Mixed-State Phases with Polarized Neutrons, 
Phys. Rev. Lett. {\bf 101}, 150404 (2008). 
\bibitem{mukunda93} N. Mukunda and R. Simon, 
Quantum Kinematic Approach to the Geometric Phase. I. General Formalism, 
Ann. Phys. (N.Y.) {\bf 228}, 205 (1993). 
\bibitem{remark} By combining an even and an odd integer, the spin state evolves into 
$(-1)^{n_{-}} \left( c_{-} \ket{-} - c_{+} \ket{+} \right)$, which is generally different from 
the input state $\ket{\psi}$. Thus, it is essential that both intergers are even or both are odd 
in order to give rise to cyclic evolution of the spin state.  
\bibitem{klein76} A. G. Klein and G. I. Opat, 
Observation of $2\pi$ rotations by Fresnel diffraction of neutrons, 
Phys. Rev. Lett. {\bf 37}, 238 (1976). 
\end{thebibliography}
\end{document}